\documentclass[%
a4paper,
prd,
twocolumn,
superscriptaddress,
preprintnumbers,
nofootinbib,
nobibnotes,
amsmath,amssymb,
aps,
floatfix
]{revtex4-2}
\usepackage{amsfonts,graphics,color}
\usepackage[english]{babel}
\usepackage{graphicx}
\usepackage{amsmath}
\usepackage{dcolumn}
\usepackage{bm}
\usepackage{multirow}
\usepackage{xcolor}
\usepackage{comment}
\definecolor{xlinkcolor}{cmyk}{1,1,0,0}
\usepackage[bookmarks=true, pdfnewwindow=true, colorlinks=true, linkcolor=xlinkcolor, citecolor=xlinkcolor, filecolor=xlinkcolor, urlcolor=xlinkcolor, final=true]{hyperref}

\usepackage[normalem]{ulem}
\definecolor{myblue}{rgb}{0.05,0.1,0.5}

\begin{document}

\preprint{INR-TH-2025-003}

\title{Tachyonic and parametric resonances for massive particle production in an intense plane wave background}

\author{Ekaterina Dmitrieva}
\email[\textbf{e-mail}: ]{edmitrieva@inr.ru}
\thanks{corresponding author}
\affiliation{Institute for Nuclear
Research of the Russian Academy of Sciences, 60th October Anniversary Prospect 7a, Moscow 117312, Russia}
\affiliation{Faculty of Physics, Moscow State University, Leninskiye Gory, 119991 Moscow, Russia}

\author{Petr Satunin}
\email[\textbf{e-mail}: ]{satunin@ms2.inr.ac.ru}
\affiliation{Institute for Nuclear
Research of the Russian Academy of Sciences, 60th October Anniversary Prospect 7a, Moscow 117312, Russia}
\affiliation{Faculty of Physics, Moscow State University, Leninskiye Gory, 119991 Moscow, Russia}

\begin{abstract}

We investigate the stability of an intensive plane wave of a massless or light field $\phi$ in a trilinear scalar model $g\phi\chi^2$ due to the resonant production of massive $\chi$ particles in a perturbatively forbidden regime. We apply two methods: first, we solve the Heisenberg equation for the quantum amplitudes of the field $\chi$ in an external plane wave background, generalizing the solution of A.Arza \cite{Arza:2020zop}. Second, for the light but massive $\phi$ we perform the relativistic boost to the rest frame of $\phi$, reducing the problem to the stability of the thoroughly investigated massive condensate. 
It turns out that the stability properties are significantly different for the cases of massless and light fields. In the first case, one should adopt the Heisenberg equation approach, since the alternative method 
cannot provide a comprehensive outcome. In the second case, the use of the Mathieu equation provides a more accurate solution, while for the massless case, this approach is not applicable.

\end{abstract}

\maketitle

\section{Introduction}

Particle production in intensive oscillating fields is an interesting phenomenon in quantum field theory (QFT) that has important applications in cosmology and astrophysics \cite{Ritus, Dittrich:2000zu, Kuznetsov:2013sea, calzetta_hu_2023, Fedotov:2022ely,grib1980quantum,Grib:1980aih}.
An interesting example of the aforementioned phenomena is the production of particles by an intensive oscillating field enhanced by parametric resonance \cite{Kofman:1997yn, lozanov2020reheating, Dufaux:2006ee, Eroncel:2025tut, WileyDeal:2025wgh}.  This process is of great interest in cosmology, since it describes extremely fast production of scalar particles driven by the oscillations of the inflaton field at the preheating stage after inflation \cite{ Kofman:1994rk, Kofman:1997yn, Traschen:1990sw, Shtanov:1994ce,  Khlebnikov:1996mc,Khlebnikov:1996wr, Allahverdi:2010xz}, see also \cite{Amin:2014eta} for a review. 
Mathematically, one solves the Heisenberg equation for the amplitudes of a certain bosonic field mode in an external oscillating field, which in the most popular two-scalar model with quartic coupling $g\phi^2\chi^2$ reduces (neglecting the expansion of the universe for simplicity) to the Mathieu equation \cite{mclachlan1947theory,jazar2021perturbation,schmitz2010reheating}. Depending on the parameters of the model, one distinguishes the analytical regimes of the 'narrow' and 'broad' resonance \cite{Kofman:1994rk,Fujisaki_1996}, which affect the preheating efficiency in different ways. Besides, the properties of the solution of this equation and the diagram of its stability are well studied mathematically and presented in many works, e.g. in \cite{mclachlan1947theory, jazar2021perturbation, abramowitz1965handbook}.  

In addition to quartic coupling, one also considers trilinear coupling $g\phi \chi^2$ \cite{Shtanov:1994ce} in which the tachyonic instability\footnote{One can regularize the theory by extra $\lambda \chi^4$ term.}  exists \cite{Andreev:1996,PhysRev.159.1089}. 
This model reads the Mathieu equation for the $\chi_k$ amplitudes as previously. 
Dufaux et al. \cite{Dufaux:2006ee} have shown that in a certain range of parameters, this instability implies a very fast tachyonic resonance for the production of particles $\chi$  from $\phi$ condensate. 

Preheating is not the only application of such a resonant particle production. Hypothetical compact objects from a Bose condensate of a scalar field (Bose stars) may have an instability due to a similar mechanism \cite{PhysRevLett.79.1607, Levkov:2020txo}.

Besides the consideration of the condensate instability due to the resonant particle production, one can also consider an intense plane wave in the initial state. 
If the plane wave is constructed from a massive field, the stability investigation reduces to the previous case of the condensate by a relativistic boost. However, it fails for a massless field, so alternative formalisms are also interesting.  One of the recently discussed examples is the instability of a plane electromagnetic wave due to the production of axions in a co-linear limit
\cite{Beyer:2021xql}.


The intense plane wave instability in a trilinear model of two scalar fields was considered in \cite{Arza:2020zop}. For a massless plane wave field, the Heisenberg equation does not imply the Mathieu equation, so a different method with additional assumptions has been applied. 
In the work \cite{Arza:2020zop}, it was shown that with a sufficiently large amplitude, a plane wave of the massless field is unstable, which leads to the formation of massive particles. For this reason, the Heisenberg equation for modes of a massive scalar field was solved. It should be noted that the calculation was carried out under the assumption of low mass of produced particles compared to the frequency of a plane wave. 

In this article, we summarize the calculations \cite{Arza:2020zop} for the case of arbitrary masses of produced particles, and also investigate in detail the connection of the method based on the Heisenberg equation with the Mathieu equation. 


The paper is organized as follows. 
In Section 2 we consider a toy model $g\phi\chi^2$ and obtain the equations of motion.
In Section 3 we study the solution of the Heisenberg equation for a high-intensity plane wave for the trilinear interaction. Solutions are considered both in the approximate case and without it.
In Section 4 we describe the intensely oscillating condensate using the Mathieu equation. In this section, we also describe the narrow, broad, and tachyon resonances.
In Section 5 we present a comparison of the two approaches for the condensate.
Section 6 is devoted to the conclusion.

\section{The Model}\label{Model}

We consider the theory of two interacting scalar fields $\phi$ and $\chi$  with a trilinear coupling described by a Lagrangian,
\begin{equation}\label{lagr}
    {\cal L} = \frac{1}{2}(\partial_\mu \phi)^2 + \frac{1}{2}(\partial_\mu \chi)^2  - \frac{1}{2}m_\phi^2 \phi^2 - \frac{1}{2}m_\chi^2 \chi^2 -g\phi\chi^2, 
\end{equation}
where $m_\phi$ and $m_\chi$ are the masses of $\phi$ and $\chi$ fields, $g$ is the trilinear coupling constant.  We are interested in studying the instability of 
the plane wave of field $\phi$
\begin{equation}
\label{Wave}
    \phi(t,x)= \Phi\cos(\omega_p t - \vec p\cdot\vec x),
\end{equation}
related to the production of $\chi$ particles.  
Here $\Phi$ is the amplitude of the plane wave, $\omega_p$ and $\vec{p}$ are its frequency and momentum. Special relativity implies,  $\omega_p^2 - \vec{p}^2 = m_\phi^2$. We consider both the cases of massless $m_\phi=0$ and the massive $m_\phi \neq 0$ field $\phi$. 

The process of $\chi$ particle production is a simple perturbative process $\phi \to \chi\chi$ if $m_\phi > 2 m_\chi$. Otherwise ($m_\phi < 2 m_\chi$), the perturbative process is forbidden due to energy-momentum conservation. However, one should take into account that the solution for a free particle $\chi$ modifies in the presence of a $\phi$ plane wave with a sufficiently large amplitude $\Phi$ which can lead to nonperturbative decay channels.


The dynamics of the fields $\chi$ and $\phi$ is governed by equations of motion (EOM),
\begin{align}
\label{eqmotion2}
&(\Box+m^{2}_\chi)\chi=-2g\phi\chi, \\
\label{eqmotion1}
&(\Box + m^2_\phi)\, \phi=-g\chi^2.
\end{align}
Eq.~(\ref{eqmotion2}) refers to the production of field $\chi$ in an external plane wave of field $\phi$, Eq.~(\ref{eqmotion1}) --- to the production of $\phi$ modes in the generated $\chi$ field. The last process can be neglected at early times while the  amplitude of the produced $\chi$ field is significantly less than those of $\phi$, and the back reaction is not taken into account. 

We perform two approaches. First, we solve equation \eqref{eqmotion2} for the amplitudes of $\chi$ field following \cite{Arza:2020zop} in the approximation used in the article and beyond it, both for massless and massive $\phi$. The second approach is related exclusively to the case of massive $\phi$ and implies the relativistic boost to the frame in which the plane wave \eqref{Wave} is a condensate,
\begin{equation}
\label{Condensate}
    \phi(t) = \Phi\cos(m_\phi t),
\end{equation}
and the problem reduces to the known Mathieu equation.

\section{The Heisenberg equation approach. Plane wave}

We solve equation \eqref{eqmotion2} with the classical field $\phi$ \eqref{Wave}. The field $\chi$ can be decomposed in Fourier series as
\begin{equation}
\label{chi-decomposition}
    \chi(t,\vec{x})=\int\frac{d^3k}{(2\pi)^3}\frac{1}{\sqrt{2\Omega_{\vec{k}}}}\Big(\chi_{\vec{k}}(t)e^{i\vec{k}\cdot\vec{x}}+\chi^*_{\vec{k}}(t)e^{-i\vec{k}\cdot\vec{x}}\Big),
\end{equation}
where $\Omega_{\vec{k}}=\sqrt{\vec k^2+m^2_\chi}$ and the operators $\chi_{\vec{k}}$ and $\chi^{\dagger}_{\vec{k}}$ satisfy the commutation relations $[\chi_{\vec{k}},\chi_{\vec{k}'}]=0$, $[\chi_{\vec{k}},\chi{\dagger}_{\vec{k}'}]=(2\pi)^3\delta^3(\vec{k}-\vec{k}')$.

Substituting the decomposition \eqref{chi-decomposition} into the equations of motion \eqref{eqmotion2}, encounting the plane wave field $\phi$ \eqref{Wave} and denoting the Bogolubov transformation  $A_{\vec{k}}=\chi_{\vec{k}}+\chi^{\dagger}_{-\vec{k}}$ one finally obtains (cf. \cite{Arza:2020zop}),
\begin{align}\label{arza20}
(\partial^2_t+\Omega^2_{\vec{k}})A_{\vec{k}}=-\omega^2_{\vec{p}}\alpha\Bigg(\sqrt{\frac{\Omega_{\vec{k}}}{\Omega_{\vec{k}-\vec{p}}}}A_{\vec{k}-\vec{p}}e^{-i\omega_{\vec{p}}t}+\\
+\sqrt{\frac{\Omega_{\vec{k}}}{\Omega_{\vec{k}+\vec{p}}}}A_{\vec{k}+\vec{p}}e^{i\omega_{\vec{p}}t}\Bigg),\nonumber
\end{align}
where $\alpha\equiv\frac{g\Phi}{\omega^2_{\vec{p}}}$. 
We highlight the standard oscillation part $\chi_{\vec{k}}$, 
$\chi_{\vec{k}}=a_{\vec{k}}(t)e^{-i\Omega_{\vec{k}}t}$ and $\chi^{\dagger}_{-\vec{k}}=a^{\dagger}_{-\vec{k}}(t)e^{i\Omega_{-\vec{k}}t}$ (where $\Omega_{\vec k}=\Omega_{-\vec k}$) and substitute it into the equation \eqref{arza20}. The time evolution for the amplitudes $a_{\vec{k}}$, $a_{-\vec{k}}^\dagger$ is governed by the equation,
\begin{multline}
\label{before_approx}
e^{-i\Omega_{\vec{k}}t}(\ddot a_{\vec k}-2i\Omega_{\vec k}\dot a_{\vec k})+e^{i\Omega_{-\vec{k}}t}(\ddot a^{\dagger}_{-\vec k}+2i\Omega_{-\vec k}\dot a^{\dagger}_{-\vec k})=\\ 
=-\omega^2_{\vec p}\alpha\Bigg(\sqrt{\frac{\Omega_{\vec k}}{\Omega_{\vec k+\vec p}}}\Big(a^{\dagger}_{-\vec k-\vec p}e^{i(\Omega_{-\vec k-\vec p}+\omega_{\vec p})t}\\ +a_{\vec k+\vec p}e^{-i(\Omega_{\vec k+\vec p}-\omega_{\vec p})t}\Big)\\ +\sqrt{\frac{\Omega_{\vec k}}{\Omega_{\vec k-\vec p}}}\Big(a^{\dagger}_{-\vec k+\vec p}e^{i(\Omega_{-\vec k+\vec p}-\omega_{\vec p})t} + a_{\vec k-\vec p}e^{-i(\Omega_{\vec k-\vec p}+\omega_{\vec p})t}\Big)\Bigg).
\end{multline}
This is an infinite system of 2-nd order differential linear equations. The right column of equation \eqref{before_approx} can be obtained from the left column using Hermitian conjugation and $\vec k\rightarrow-\vec k$ substitution. We can conditionally separate equation \eqref{before_approx} into two parts, one part corresponds to $a_{\vec k}$, and the other part corresponds to $a^\dagger_{-\vec k}$. Consequently, the solutions for the amplitudes of each of these parts will contain parameters associated with $a_{\vec k}$ or $a^\dagger_{-\vec k}$, respectively. Terms with $a^{\dagger}_{\vec p-\vec k}$ and $a^{\dagger}_{-\vec p-\vec k}$ correspond momentum $\vec k$ and terms with $a_{\vec k+\vec p}$ and $a_{\vec k-\vec p}$ correspond momentum $-\vec k$. 
The general solution is a linear combination of complex and real exponents. If the solution contain exponential growing with time term,
$a_{\vec{k}}(t)=\dots + A_{\vec{k}'}\mbox{e}^{c_{\vec{k}'}\,t} a^\dagger_{\vec{k}'}(0) + \dots$ 
for certain $\vec{k}'$, the occupation number for modes of momentum $\Vec{k}'$ grows resonantly,
$$
    f_{\chi,\vec{k}'}(t)=\langle0 |a^\dagger_{\vec k}(t) a_{\vec k}(t)| 0\rangle=
A^2_{\vec{k}'}\mbox{e}^{2c_{\vec{k}'}\,t} + ... 
$$
In case of several resonant terms one should sum over all of them.

  Note that the terms in the r.h.s. of any line of eq.~(\ref{before_approx}) are obtained from the corresponding l.h.s. terms of that line. Thus, it is sufficient to search for the resonant solution in the l.h.s. column of  eq.~(\ref{before_approx}) separately.
Denoting additionally
\begin{equation}
\sigma_{\vec{p}-\vec{k}}=-\omega^2_{\vec p}\alpha\sqrt{\frac{\Omega_{\vec k}}{\Omega_{\vec p-\vec k}}} \ \ \mbox{and}\ \ \sigma_{\vec{p}+\vec{k}}=-\omega^2_{\vec p}\alpha\sqrt{\frac{\Omega_{\vec k}}{\Omega_{\vec p+\vec k}}},
\end{equation}
we rewrite eq.~(\ref{before_approx}) as 
\begin{multline}
  e^{-i\Omega_{\vec{k}}t}(\ddot a_{\vec k}-2i\Omega_{\vec k}\dot a_{\vec k})=\sigma_{\vec{p}-\vec{k}}a^{\dagger}_{\vec p-\vec k}e^{i(\Omega_{\vec p-\vec k}-\omega_{\vec p})t} +\\ \sigma_{\vec{p}+\vec{k}}a^{\dagger}_{-\vec p-\vec k}e^{i(\Omega_{-\vec p-\vec k}+\omega_{\vec p})t}.
  \label{amp1}
\end{multline}
The term with $a^{\dagger}_{\vec p-\vec k}$ is related to the production of $\chi$ modes with momentum $\Vec{p}-\vec{k}$, $a^{\dagger}_{-\vec p-\vec k}$ --- with momentum  $(-\vec{p}-\vec{k})$. 

\subsection{The solution using approximation}\label{sec_approx}

The equation for the amplitude (\ref{amp1}) reads (amplitude $a^{\dagger}_{\vec p-\vec k}$ is leading),
\begin{equation}\label{eq_p-k}
\ddot a_{\vec k}-2i\Omega_{\vec k}\dot a_{\vec k}=\sigma_{\vec{p}-\vec{k}}a^{\dagger}_{\vec p -\vec k}e^{i(\Omega_{\vec{k}} + \Omega_{\vec p-\vec k}-\omega_{\vec p})t}.
\end{equation}
This equation can be simplified if $a_{\vec{k}}(t)$ varies slower with time than $\chi_{\vec{k}}(t)$ in a way that we can neglect the second derivative in (\ref{eq_p-k}), 
\begin{equation}
\label{RWA}
|\ddot{a}_{\vec{k}}| \ll |\Omega_{\vec{k}} \dot{a}_{\vec{k}} |,
\end{equation}
which limit is referred to as 
the rotating wave approximation (RWA) in \cite{Arza:2020zop}. Thus, in this approximation the first term in (\ref{eq_p-k}) can be neglected:
\begin{equation}
    \dot a_{\vec k}=-i\frac{\sigma_{\vec p-\vec k}}{\Omega_{\vec k}}a^{\dagger}_{\vec p-\vec k}e^{i\epsilon_{\vec k} t}.
\end{equation}
The solution \cite{Arza:2020zop}  reads (in our notations), 
\begin{multline}\label{sol0}
a_{\vec{k}}(t)=e^{i\epsilon_{\vec k}t/2}\Bigg(a_{\vec k}(0)\Big(\cosh(s^0_{\vec k}t)-i\frac{\epsilon_{\vec k}}{2s^0_{\vec k}}\sinh(s^0_{\vec k}t)\Big)\\-
i\frac{\sigma_{\vec p-\vec k}}{2s^0_{\vec k}\Omega_{\vec k}}a^{\dagger}_{\vec p-\vec k}(0)\sinh(s^0_{\vec k}t)\Bigg),
\end{multline}
where
\begin{equation}
\label{sk0}
    s^0_{\vec k}=\frac{1}{2}\sqrt{\frac{\sigma^2_{\vec{p}-\vec{k}}}{\Omega_{\vec{k}}^2} -\epsilon^2_{\vec{k}}}, \qquad  \epsilon_{\vec{k}} = \epsilon_{\vec{p}-\vec{k}}=\Omega_{\vec{k}}+\Omega_{\vec{p}-\vec{k}}-\omega_{\vec{p}}.
\end{equation}
In order to obtain the resonance the argument of the hyperbolic sine in the last term of (\ref{sol0}) should be real, Im $s^0_{\vec k} = 0$.  Note that there is a symmetry regarding the replacement of $\vec k$ by $\vec p-\vec k$. This is in agreement with the production of two particles in the final state.
For fixed $\alpha, m_\chi$, the momentum threshold for resonance \eqref{sol0},\eqref{sk0} reads $k=p/2$. 

Now we return to the limits of applicability of this result. Substituting the solution (\ref{sol0}),(\ref{sk0}) into the  approximation (\ref{RWA}), we obtain: 
\begin{equation}
\epsilon_{\vec{k}} \ll  \Omega_{\vec k}, \qquad s^0_{\vec k} \ll \Omega_{\vec k}.
\end{equation}
The first condition reads, $\Omega_{\vec{p}-\vec{k}} \ll \omega_{\vec{p}}$, which restricts us to the cases of small  $m_\chi \ll \omega_{\vec p}$\footnote{More accurately, $m_\chi^2- \frac{m_\phi^2}{4} \ll \omega_p^2$.} and not very small  $|\vec k | \gg \frac{m_\chi^2}{2\omega_{\vec p}}$. The second condition reduces to $\alpha \ll 1$. 


The occupancy number for a mode of a field $\chi$ with momentum $\vec k$ grows resonantly \cite{grib1980quantum}
\begin{equation}\label{eq_number}
    f_{\chi,\vec{k}}(t)=\langle0 |a^\dagger_{\vec k}(t) a_{\vec k}(t)| 0\rangle=
    \frac{\sigma^2_{\vec p-\vec k}}{4}\frac{\sinh^2(s^0_{\vec k}t)}{\left(s^0_{\vec k}\right)^2\Omega_{\vec k}^2}
\end{equation}
The total density of all resonantly produced modes reads,
\begin{equation}
\label{density}
    n_\chi (t) = \int \frac{d^3k}{(2\pi)^3} f_{\chi,\vec{k}}(t),
\end{equation}
where the momentum integral $\int d^3 k$ should be taken only over momenta satisfying the resonant condition $\left(s^{0}_{\vec k}\right)^2 > 0$. 
Consequently, the condition of the resonance boundary reads,
\begin{equation}
\label{rcond}
    s_k^0 =0. 
\end{equation}
The resonance condition \eqref{rcond}, \eqref{sk0} for the threshold value $k=p/2$ is 
\begin{equation}
  \frac{4g^2\Phi^2}{\omega_p^2 + 4 m_\chi^2 - m_\phi^2} = \left( \sqrt{\omega_p^2+4 m_\chi^2 - m_\phi^2} - \omega_p\right)^2.  
\end{equation}
In the limit of our approximation $m_\chi \ll \omega_p$, one derives 
that the resonance condition does not depend on $\omega_p$,
\begin{equation}
\label{eq-boundary-gen}
    g\Phi = m_\chi^2 - \frac{m_\phi^2}{4}.
\end{equation}
In the case of massless $\phi$ (as well as in the case $m_\phi \ll m_\chi$) one divides each side of eq.~\eqref{eq-boundary-gen} by $\omega_p^2$ obtaining a dimensionless bound (cf.\cite{Arza:2020zop}),
\begin{equation}\label{bound_arza}
    \alpha = \mu^2_\chi, \qquad \mu^2_\chi \equiv \frac{m_\chi^2}{\omega_p^2},
\end{equation}
while in the case of nonzero $m_\phi$ one may rewrite \eqref{eq-boundary-gen} using relativistic invariant dimensionless parameters,
\begin{equation}
\label{mphi-bound}
    \frac{4g\Phi}{m_\phi^2} = \frac{4m_\chi^2}{m_\phi^2} - 1.
\end{equation}  

\subsection{The solution without approximation}\label{sec_solut_without_approx}

Here we are looking for a solution which covers arbitrary $m_\chi$ (not only $m_\chi \ll \omega_{\vec p}$ ). For this reason, we solve the equation \eqref{eq_p-k} without an approximation.



After including the phase $\epsilon_{\vec k}t/2$ into the definition of the amplitude operator, we obtain
\begin{equation}
\label{ab}
    a_{\vec{k}}=b_{\vec{k}}e^{i\epsilon_{\vec{k}}t/2}.
\end{equation}
In terms of $b_{\vec k},\; b^\dagger_{\vec k}$ operators we rewrite eq.~(\ref{eq_p-k}) as
\begin{equation}
\label{beq}
\ddot b_{\vec{k}}-i\left(2\Omega_{\vec k}-\epsilon_{\vec{k}}\right)\dot b_{\vec{k}}+\left(\Omega_{\vec{k}}\epsilon_{\vec{k}} - \frac{\epsilon^2_{\vec{k}}}{4}\right)b_{\vec{k}}=\sigma_{\vec{p}-\vec{k}}b^{\dagger}_{\vec{p}-\vec{k}}.
\end{equation}
We determine the solution of eqs.~(\ref{beq}),(\ref{ab}) using an ansatz of the form of eq.~(\ref{sol0}) with different values of coefficients. Precisely, the solution reads,
\begin{multline}
\label{solution}
a_{\vec{k}}(t)=e^{i\epsilon_{\vec{k}}t/2}\Bigg[  a_{\vec{k}}\,(0)\Big(\cosh(s_{\vec k}t) \\ \;-\; i\frac{\epsilon^2_{\vec{k}}/4-s^2_{\vec k}-\Omega_{\vec{k}}\epsilon_{\vec{k}}}{s_{\vec k}(2\Omega_{\vec{k}} -\epsilon_{\vec{k}})}\sinh(s_{\vec k}t)\Big)\;-  \\ 
\;-\;  a^{\dagger}_{\vec{p}-\vec{k}}(0)\;\cdot \;i\frac{\sigma_{\vec{p}-\vec{k}}}{s_{\vec k}(2\Omega_{\vec{k}} - \epsilon_{\vec{k}})}\sinh(s_{\vec k}t)\Bigg],
\end{multline}
where
\begin{equation}\label{eq:s}
s^2_{\vec k}=\sqrt{\Omega^2_{\vec k}(2\Omega_{\vec k}-\epsilon_{\vec k})^2+\alpha^2\omega^4_p\frac{\Omega_{\vec k}}{\Omega_{\vec{p}-\vec k}}
}-\Omega_{\vec k}(2\Omega_{\vec k}-\epsilon_{\vec k})-\frac{\epsilon_{\vec k}^2}{4}.
\end{equation}
Expanding the expression ~\eqref{eq:s} into the Taylor series over a small parameter $\alpha \ll 1$ and additionally over $\epsilon_{\vec{k}}/\Omega_{\vec{k}} \ll 1$, one derives the approximated result eq.~\eqref{sk0} in the leading order.

 The result \eqref{eq:s} is not invariant under symmetry $\vec k\rightarrow \vec p-\vec k$ beyond the leading-order perturbative expansion~\eqref{sk0}. 
 Thus, in general \eqref{solution}, \eqref{eq:s} is still a solution only if $\vec k=\frac{\vec p}{2}$. This may indicate that the ansatz we have chosen is incomplete.

Note that if we do not take into account the absence of symmetry, then $k = p/2$ (or $\kappa=0.5$ in dimensionless case) still remains the lower, and therefore the most general, boundary of instability.


\begin{figure}[h]
      \centering
      \includegraphics[width=0.99\linewidth]{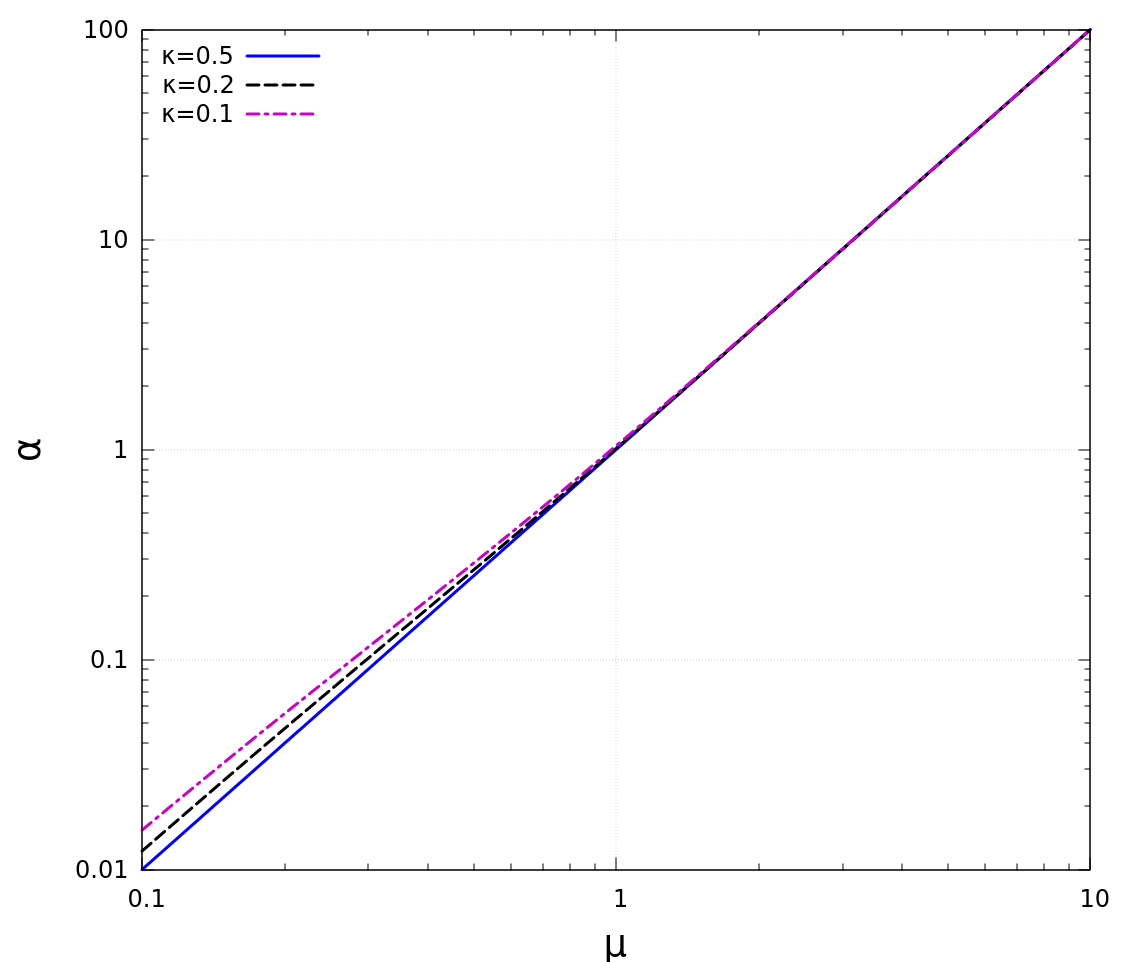}
      \caption{Dependence $\alpha_{\vec \kappa}(\mu_\chi)$ at fixed $\kappa=0.1, 0.2, 0.5$. The upper region related to the resonance. }\label{alp(mu)}
\end{figure}

The occupancy number for a mode of a field $\chi$ with momentum $\vec k$ grows resonantly \cite{grib1980quantum},

\begin{equation}\label{f_no_approx}
    f_{\chi,\vec{k}}(t)=\langle0 |a^\dagger_{\vec k}(t) a_{\vec k}(t)| 0\rangle=
    \frac{\sigma^2_{\vec p-\vec k}}{4}\frac{\sinh^2(s_{\vec k}t)}{s^2_{\vec k}\epsilon^2_{\vec k}}.
\end{equation}

The total density of all resonantly produced modes is calculated using the same formula \eqref{density} as in the previous case. However, the resonance condition is determined by $s^{2}_{\vec k} > 0$.

In order to precisely determine the integration domain, one should solve $s^{2}_{\vec k} = 0$ for (\ref{eq:s}) with respect to the momentum for various numbers of parameters.

In addition, it is convenient to introduce dimensionless notation
$\vec{\kappa}=\vec{k}/\omega, \; \mu_\chi=m_\chi/\omega,\; \vec{v}=\vec{p}/\omega$. The dimensionless frequencies reads, 
\begin{eqnarray}
\label{beta}
\frac{\Omega_{\vec k}}{\omega_{\vec p}} \equiv  \beta_{\vec{\kappa}}=\sqrt{\kappa^2 +\mu^2_\chi}, \qquad\\ \frac{\Omega_{\vec p - \vec k}}{\omega_{\vec p}} \equiv  \beta_{\vec v - \vec{\kappa}}=\sqrt{1  -2\kappa  + \kappa^2 +\mu^2_\chi}.\nonumber
\end{eqnarray}
In terms of dimensionless variables (\ref{beta})  eq. (\ref{eq:s}) reads (furthermore, we introduce the dimensionless parameter $\eta_{\vec k} = s_{\vec k}/\omega_{\vec p}$),

\begin{multline}\label{eta}
  \eta^2_{\vec k}=\sqrt{\beta_{\vec{\kappa}}^2\left(\beta_{\vec{\kappa}}-\beta_{\vec{v}-\vec{k}}+1 \right)^2 + \alpha^2\beta_{\vec{\kappa}}\beta_{\vec{v}-\vec{\kappa}}^{-1}} \\- \beta_{\vec{\kappa}}\left(\beta_{\vec{\kappa}}-\beta_{\vec{v}-\vec{k}}+1 \right) -\frac{(\beta_{\vec{\kappa}}+\beta_{\vec{v}-\vec{k}}-1)^2}{4}.  
\end{multline}

Fixing $\vec \kappa$ in the equation (\ref{eta}), we determine the threshold value of $\alpha$ for fixed $\vec \kappa$, 
\begin{align}\label{alp}
\alpha_{\vec \kappa}=\frac{1}{4}\sqrt{\frac{\beta_{\vec{v}-\vec{\kappa}}}{\beta_{\vec{\kappa}}}(\beta_{\vec{\kappa}}+\beta_{\vec{v}-\vec{\kappa}}-1)^2(-3\beta_{\vec{\kappa}}+\beta_{\vec{v}-\vec{\kappa}}-1)^2}.
\end{align}

This dependence is shown in Figure \ref{alp(mu)} for several values of momentum $\vec k$.


The resonance condition eq.~\eqref{eq:s} at the threshold $k=p/2$ reads directly eq.~\eqref{eq-boundary-gen} without any additional approximation on $\omega_p$. This illustrates the fact that the value of $\omega_p$ determines only the frame; the stability boundary does not depend on it due to the Lorentz invariance.

\subsection{Problems of the Heisenberg equation approach}

The \eqref{solution}-\eqref{eq:s} are not symmetric with respect to the substitution $\vec k~\longrightarrow~\vec p-\vec k$ unless $\vec k=\vec p/2$. 
In the approximate case $\alpha \ll 1$, the symmetry is restored.  

This result may indicate that the ansatz \eqref{solution} is incomplete and other research methods are required.  
Hence, we consider an alternative method in the next section.

\section{Mathieu equation analysis for the instability in condenstae}





In this section, we consider a massive plane wave that has undergone the Lorentz boost (see detailed in Appendix \ref{app_boost}) to its rest frame being a condensate \eqref{Condensate}; $\omega_p = m_\phi, \ \vec{p}=0$. The amplitude $\Phi$ is a Lorentz scalar, so it has not changed after the boost.

The EOM for the mode $\chi_{\vec k}$ reads,
\begin{equation}
\label{ddot1}
    \Ddot{\chi}_{\vec k}  + \left(k^2 + m_\chi^2 + 2g\Phi \cos(m_\phi t) \right) \chi_{\vec k} = 0,
\end{equation}
which is the Mathieu equation,
\begin{equation}
\label{Matthieu}
    \chi_{\vec k}'' + \left( A_k + 2q \cos(2z)\right)\chi_{\vec k} = 0,
\end{equation}
where $z=\frac{m_\phi t}{2}$, and
\begin{equation}
\label{Aq}
    A_k = 4\frac{k^2+m_\chi^2}{m_\phi^2}, \qquad q = 4\frac{g\Phi}{m_\phi^2}.
\end{equation}
The number of modes is determined as,
\begin{equation}
    n_{\vec k} = \langle 0| a_{\vec k}(t) a^\dagger_{\vec k}(t) |0\rangle = |\chi_{\vec k} (t)|^2 
\end{equation}

\begin{figure}
    \centering
    \includegraphics[width=0.99\linewidth]{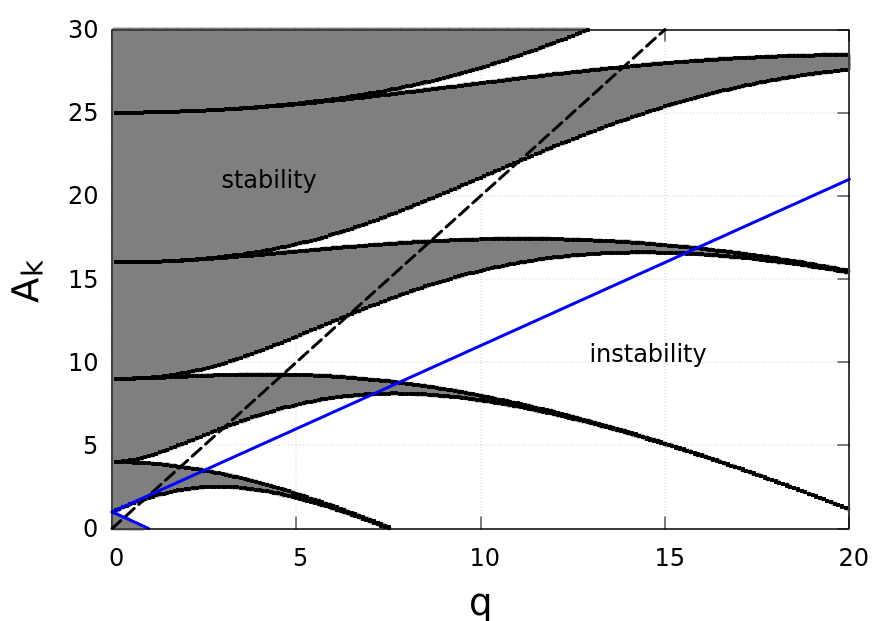}
    \caption{Stability diagram of the Mathieu equation and stability bounds for the Heisenberg case (equation \eqref{eq_q_w_app})
    . Vertical axis: $A_k$, Horizontal: $q$. Grey area: stability regions of Mathieu equation. Blue solid line: Heisenberg instability bound, see eq.~\eqref{eq_q_w_app}. 
    Black dashed line - line $A_k=2q$.}
    \label{fig_Mathieu_stab/inst_Arza}
\end{figure}

The Mathieu equation and its solutions are extensively researched in the literature \cite{mclachlan1947theory,jazar2021perturbation}. It appears in physical systems which undergo parametric resonance, such as a mathematical pendulum in an external driving force. In the parametric plane $(A_k,q)$ one distinguishes the area of stability, where the solutions of the Mathieu equation oscillate, and the area of instability, where one of the solutions exponentially grows. This stability chart is shown in Fig.\ref{fig_Mathieu_stab/inst_Arza}, where the gray area refers to stability, while the white area - to the instability.

Besides the pendulum in classical mechanics,  the Mathieu equation for given modes in field theory arises in the context of resonant matter particle production by an inflaton field at the end of inflation (preheating), \cite{Kofman:1997yn}, see \cite{schmitz2010reheating} for review. The Mathieu equation appears in the most studied model $g\phi^2\chi^2$ which does not have kinematical instability, 
and in the trilinear model $g\phi \chi^2$. In inflation models, the mass of the matter particles is usually significantly less than the mass of the inflaton, $m_\chi \ll m_\phi$, so the $m_\chi^2$ term in \eqref{ddot1} is usually neglected, taking $A_k = \frac{4k^2}{m_\phi^2}$. Otherwise, in our task $m_\chi > m_\phi /2$, we cannot neglect it. The case $k=0$ related to the threshold $k=p/2$ in the unboosted case.

Depending on the model parameters, one distinguishes the analytically solvable cases of narrow and broad resonances.


\begin{figure}
    \centering
    \includegraphics[width=0.48\linewidth]{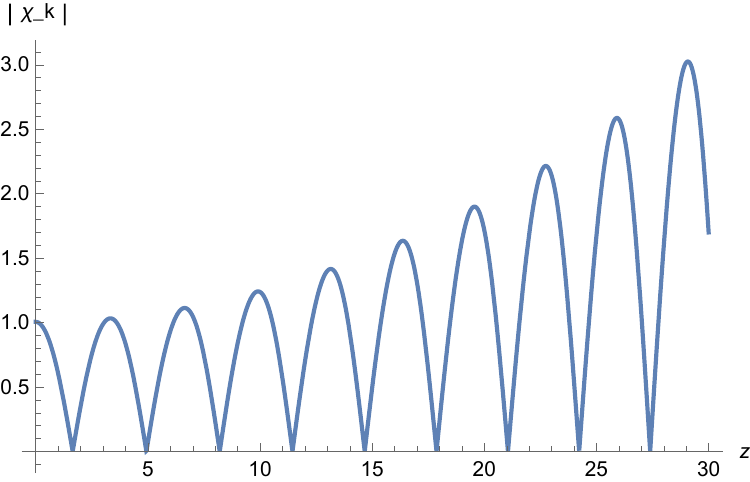}
    \includegraphics[width=0.499\linewidth]{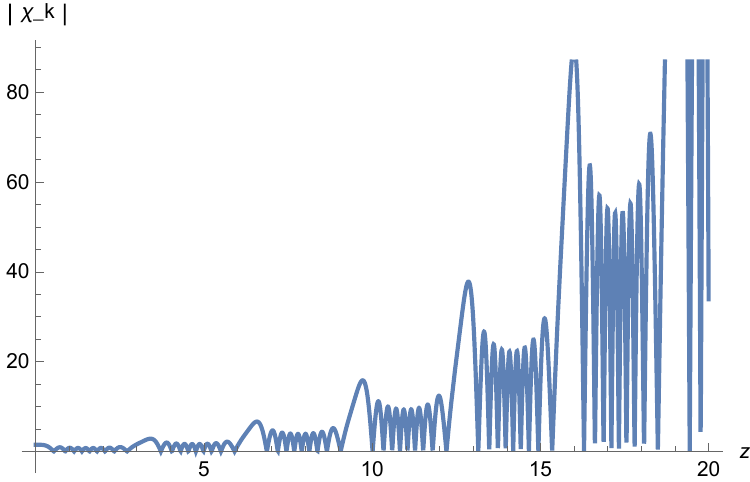}\\
    \includegraphics[width=0.3\linewidth]{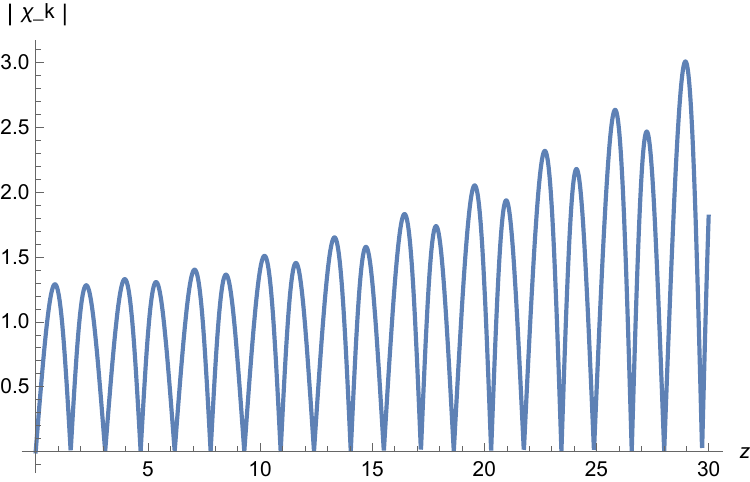}
    \includegraphics[width=0.3\linewidth]{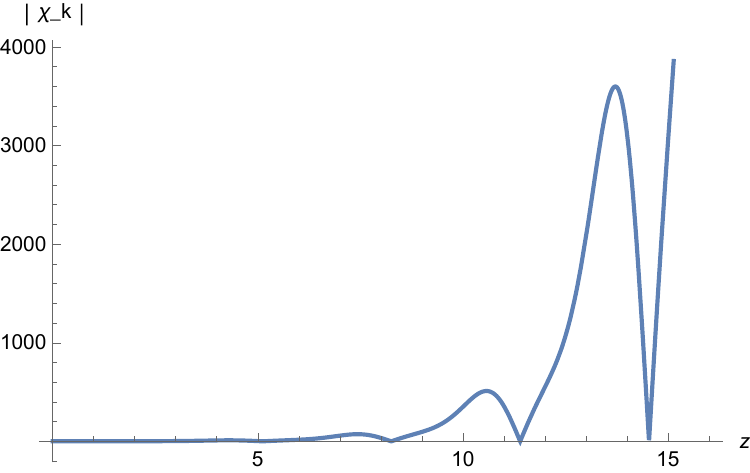}
    \includegraphics[width=0.3\linewidth]{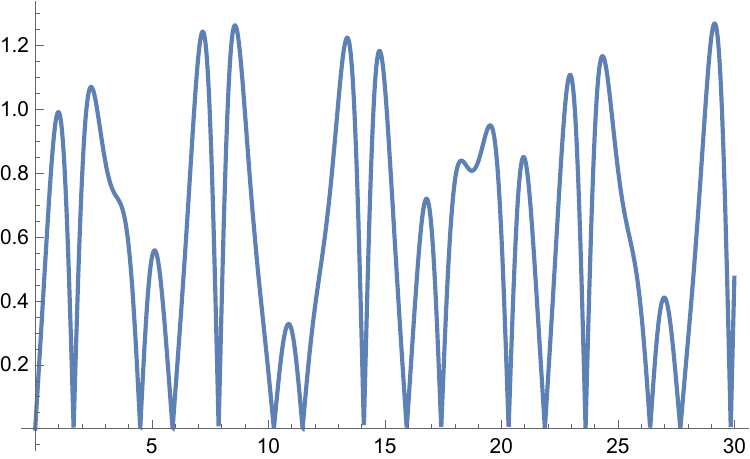}
    \caption{The resonant solutions. Upper line. Left panel: Narrow resonance, $A_k=1$, $q=0.1$. Right panel: Broad resonance. $A_k=100$, $q=50$. Lower line. Left panel: $N=2$ band resonance. $A_k=4$, $q=1$. Central panel: Tachyonic resonance.  $A_k=1$, $q=1.5$. Right panel: destructive parametrical resonance in naively tachyonic unstable area.  $A_k=3$, $q=2$}
    \label{fig:NarrowBroad-label}
\end{figure}

\subsection{Narrow resonance}

The regime $q \ll 1$ shows a simple solution in terms of elementary functions \cite{landau1982mechanics}. The growing solution $\chi_{\vec{k}}$ of eq.~\eqref{Matthieu} can be considered as an oscillating function modulated by a growing exponent, 
\begin{equation}
    \chi_{\vec{k}} (z) = e^{\mu z} \cos(z + \varphi),
\end{equation}
where $\varphi$ just defines 
the initial condition, and the exponent $\mu$ reads \cite{landau1982mechanics},
\begin{equation}
\label{Narrow}
  \mu=\frac{1}{2}\sqrt{q^2-(A_k-1)^2}.
\end{equation}
A graph illustrating a narrow resonant solution is shown in the up-left corner of Fig.~\ref{fig:NarrowBroad-label}. The number of $\chi$ particles grows exponentially, $n_\chi = |\chi_{\vec{k}}|^2 \sim e^{2\mu z}$ which determines the so-called narrow resonance \cite{Kofman:1997yn, lozanov2020reheating, mclachlan1947theory}. 
The condition $\mu=0$ shows the boundary of the resonance,
\begin{equation}\label{bound_first_peak}
    q = |A_k - 1|.
\end{equation}
This dependence is shown graphically in the lower left corner of Figure \ref{fig_Mathieu_stab/inst_Arza} (blue line). 
Using the notation \eqref{Aq} with $m_\chi > m_\phi/2$, and setting the threshold zero momentum $k=0$, one obtains the instability bound \eqref{mphi-bound}.

There is the following interpretation for this phenomenon.
We take one massive particle $\phi$ from the condensate, and consider decay $\phi \to 2\chi$. 
Due to the fact that $\chi$ particles are bosons, their presence in the final state causes Bose enhancement, which ultimately leads to an exponential increase in the number of $\chi$ particles produced. This process is the main resonance effect for the narrow peak\cite{Kofman:1997yn, Allahverdi:1999je, lozanov2020reheating}.

Several massive particles of $\phi$ can merge producing $\chi$ particles, $N\phi \to 2\chi$, with subsequent  Bose enhancement. The corresponding solutions (characteristic functions of Mathieu \cite{mclachlan1947theory, jazar2021perturbation}) of eq.~\eqref{Matthieu} give additional narrow peaks at Fig.2. that reach $q=0$ at $A_k = N^2$, each of those narrower than the previous.




The solution for $N=2$ is shown in the lower left part of Figure \ref{fig:NarrowBroad-label}. We see that there is an effect of additional oscillations called beats. 
In large $N \gg 1$ this supplementary oscillation takes the form seen in Fig.~\ref{fig:NarrowBroad-label} right panel. 

Note that the additional narrow peaks ($N > 1$) of the Mathieu equation, after boost transformation, lead to the effect of $\chi$ particle production in an interaction of several quanta from a massive plane wave $\phi$. In the massless limit, the effect vanishes since two or more quanta of massless field propagate with the same velocity and do not interact with each other. Thus, the instability boundary for the massless case does not include these $N > 1$ peaks. 

\subsection{Broad resonance}\label{broad_res}

The narrow resonance analytical solution fails in the regime $q \gtrsim 1$. However, there is another analytical solution called broad resonance in the regime $g \gg 1$ which is even more efficient.

The broad resonance regime is defined so that during one period of $\phi$ field oscillation, the field $\chi$ oscillates many times. In this case, the frequency of the $\chi$ field (equation \eqref{ddot1}) is determined as follows,
\begin{equation} \label{omega_chi}
\omega=\sqrt{k^2+m^2_\chi+2g\phi(t)}.
\end{equation}
 The graph of the solution corresponding to a broad parametric resonance is shown in the upper right corner of Figure \ref{fig:NarrowBroad-label}.    

In the end of inflation, preheating can start with a broad resonance and continue as a narrow \cite{Kofman:1997yn}. During a broad resonance, particles are produced by bursts, which are divided in time by $\sim T$, where $T = \frac{2\pi}{m_\phi}$ is the period of oscillation of the condensate. These bursts occur every time when the condition of adiabaticity \cite{lozanov2020reheating,Dufaux:2006ee} 
\begin{equation}\label{cond_adiab}
\frac{\dot\omega(k,t)}{\omega^2(k,t)}\ll1,\end{equation}
is violated.
Inserting \eqref{omega_chi} into \eqref{cond_adiab} yields the broad resonance condition for the Mathieu equation 
\begin{equation}\label{cond_adiab_Mathieu}
2q\sin(2z)\gtrsim(A_k+2q\cos(2z))^{\frac{2}{3}}.\end{equation}
This expression implies two different conditions for the broad resonance \cite{lozanov2020reheating}

\begin{equation}\label{cond_broad1}
    A_k<2|q|, \quad  m_\phi t=\pi/2+\pi N,
\end{equation}
or

\begin{equation}\label{cond_broad2}
    A_k-2|q|\ll|q|^{1/2}, \quad m_\phi t=\pi+2\pi N.
\end{equation}

In the laboratory frame, particle production due to the broad resonance occurs by bursts as well, each burst arising at the beginning of the period $T=2\pi/\omega_p$. However, in the theory with massless $\phi$  there is no rest frame. Moreover,  both conditions \eqref{cond_broad1}, \eqref{cond_broad2} are not satisfied in the limit $m_\phi \to 0$.  Consequently, a broad resonance does not emerge in the massless case.


\paragraph{Tachionic instability}
The parameters of the tachyon instability satisfy the condition of broad resonance, but there is an additional requirement.
For a trilinear interaction at $q>A_k/2$ the eq.(\ref{Matthieu}) shows that that for several ranges of z the effective mass becomes negative (tachyonic) for that range of z, the following solution is expected to exponentially increase in that range, so the instability under the line $A_k=2q$ in Fig.\ref{fig_Mathieu_stab/inst_Arza} is called the tachyon instability \cite{Dufaux:2006ee, lozanov2020reheating}. Tachyon preheating can be much more efficient than a conventional preheating mechanism. It is important to consider this phenomenon in the context of our work, as previously such a case of the ratio of the mass of condensate and the particles produced ($m_\phi<2m_\chi$) had not been investigated in the literature.

If we consider only positive $q$, then the conditions of tachyon resonance will look like this
\begin{equation}\label{cond_tach_res}
    0<A_k<2q.
\end{equation}

The conditions for broad resonance, which were described in the previous section, show that the tachyon resonance region is part of the broad resonance region \cite{lozanov2020reheating}. 

Comparative analysis of the conditions of tachyon and broad resonances demonstrated that a range of parameters in which broad resonance will be observed, but there will be no tachyon resonance.
We will examine the second condition for broad resonance and, at the same time, the condition that tachyon resonance does not occur:
\[A_k-2q\ll q^{1/2}\ \mbox{and} \ A_k>2q.\]
Consider $A_k=2q+\varepsilon$, where $\varepsilon$ is some small parameter, then 
$2q+\varepsilon-2q\ll q^{1/2}\Rightarrow~\varepsilon\ll q^{1/2}~\Rightarrow~\varepsilon\ll\frac{\sqrt{2g\Phi}}{m_\phi}$.
From this, we can conclude that we converge to the parameter space of broad resonance and remain outside the region of tachyon instability when deviating from $A_k=2q$ by an $\varepsilon$ value.

On the basis of the results of this paragraph, it can be concluded that tachyon resonance satisfies the conditions of broad resonance, but it has an additional constraint. As a result, it turns out that for a massless plane wave, the solution mainly belongs to 
the region of tachyon instability. It also turns out that there is only a small range of parameters in which the latter is not observed, but there is an exceptionally broad resonance.

\section{Comparison of two approaches for condensate}

In this section, we show that the Heisenberg equation solution for the decay of a massive condensate corresponds exactly to 
a narrow resonance, and compare it with the exact Matheiu chart.
\paragraph{The approach through the Heisenberg equation for condensate}

The equation \eqref{eq_p-k} for mass $m_\phi\neq0$ and momentum $p=0$ reads,
\begin{align}
    e^{-i\Omega_{\vec k}t}\big[\ddot a_{\vec k} -2i\Omega_{\vec k}\dot a_{\vec k}+2\alpha\omega^2\cos(m_\phi t)a_{\vec k}\big]+\\+e^{i\Omega_{\vec k}t}\big[\ddot a^{\dagger}_{-\vec k}+2i\Omega_{\vec k}\dot a^{\dagger}_{-\vec k}+2\alpha\omega^2\cos(m_\phi t)a^{\dagger}_{-\vec k}\big]=0
\end{align}
Including the rotation $a_{\vec k}(t)=b_{\vec k}(t)e^{i\epsilon_{\vec k} t/2}$, we obtain the equation for $b_{\vec k}$
\begin{equation}\label{eq_w_app}
    \ddot b_{\vec k}+i\dot b_{\vec k}(\epsilon_{\vec k}-2\Omega_{\vec k})+b_{\vec k}(\epsilon_{\vec k}\Omega_{\vec k}-\frac{\epsilon^2_{\vec k}}{4})+\alpha m_\phi^2 b^{\dagger}_{-\vec k}=0.
\end{equation}
We replace
an analogue of the approximate solution type. Naturally, our solution, when substituting parameters that satisfy the approximate result,
reduces to the expression
in Section \ref{sec_solut_without_approx}.
The solution to equation \eqref{eq_w_app} is
\begin{align}\label{solut_Arza_M}
 \!\!\!\!   b_{\vec k}(t)=&b_{\vec k}(0)\!\left(\!\cosh(st)\!-\! i\frac{\frac{\epsilon_{\vec k}^2}{4}-s^2-\Omega_{\vec k}\epsilon_{\vec k}}{s(2\Omega_{\vec k}-\epsilon_{\vec k})}\sinh(st)\!\right)\!-\!\notag\\-i&b^{\dagger}_{-\vec k}\frac{\alpha\omega^2}{s(2\Omega_{\vec k}-\epsilon_{\vec k})}\sinh(st)\!,
\end{align}
where
\begin{equation}
s=\sqrt{\Omega^2_{\vec k}(2\Omega_{\vec k}-\epsilon_{\vec k})^2+\alpha^2 m_\phi^4}-\Omega_{\vec k}(2\Omega_{\vec k}-\epsilon_{\vec k})-\frac{\epsilon^2_{\vec k}}{4}.
\end{equation}
In terms of the parameters of the Mathieu equation, we obtain the following relation for the stability boundary $s^2=0$ for any $k$:
\begin{equation}\label{eq_q_w_app}
    q=|A_k-1|.
\end{equation}
This is explicitly the instability bound for the narrow resonance.
This expression is completely consistent with equation \eqref{bound_first_peak} for the first peak of the Mathieu instability boundary. It is depicted in Figure \ref{fig_Mathieu_stab/inst_Arza} by the solid blue line.

These results demonstrate that the ansatz used describes a narrow resonance and only the first peak, even for the massive case. Consequently, in the case $m_\phi \neq 0$, one should use a more complete ansatz or conduct the study using the Mathieu equation.

\section{Conclusion}

In this article, we examined the process of massive particle production in a plane wave background of an intensive massless or  light
 field  in a toy scalar model with interaction $g\phi\chi^2$. We applied two methods: the solution of the Heisenberg equation for the amplitudes, and reduction to the known case of particle production in a massive condensate, which is mathematically described by the Mathieu equation, and compared them. In addition, we compare both methods for the condensate.   

The Heisenberg equation method provides a solution \cite{Arza:2020zop} which has the form of the narrow resonance in a massive scalar condensate, but works also well in a different physical model, 
related  to the tachyon instability. 

The author of \cite{Arza:2020zop} presents the method in the case of low mass; we considered the case of an arbitrary mass. We have found an analytical solution to the Heisenberg equation for the amplitude of the $\chi$ field for arbitrary model parameters. We also showed that resonance is possible beyond the approximation discussed in \cite{Arza:2020zop}. Subsequently, we calculated that in the case of a large mass (here we considered the case when the initial plane wave $\phi$ is massless or has a mass less than the mass of two $\chi$ particles), a higher value of the amplitude is required to achieve resonance. 
This implies the existence of a threshold value for the plane wave amplitude, which is given by the inequality
\[\Phi \geq \frac{m_\chi^2}{g}.\] 

The second solution method is better suited to describie massive plane waves. We reduce the equation of motion for a low-mass condensate to the Mathieu equation and analyze the instability diagram for the resulting parameters. Note that for the massless case, the parameters correspond to the broad resonance regime, but no resonance actually occurs in the limit $m_\phi\rightarrow0$. Consequently, this method cannot be applied in such cases.
We obtained a solution to the Heisenberg equation for a light condensate and compared it with the previously obtained solution of the Mathieu equation for the same physical scenario. By analyzing the instability boundary derived from the Heisenberg equation solution and comparing it with the Mathieu stability diagram, we observed that the first solution aligns with the first peak of the stability diagram within a narrow resonant region. Moreover, this comparison showed that the solution of the Heisenberg equation describes the tachyon instability, except for a constrained region of light $\chi$.
However, for systems with significant mass, the Heisenberg equation-based approach proves inadequate. This is because it provides only an approximate description of the first peak in a narrow resonance regime and fails to accurately characterize the parameter space associated with a broad resonance. Among other findings, we note that for a heavy $\chi$ particle ($m_\chi \gg m_\phi$) the instability threshold is two non-relativistic particles with zero momentum $k=0$. 
The corresponding $A_{k=0} \gg 1$ is related to the resonance, which is not narrow but broad/tachyonic.

On the other hand, when we considered a massless plane wave and its instability, we found that in this case there would be only one line of instability. Based on the particle interpretation ($n\phi\rightarrow2\chi$), for the second and subsequent peaks, two or more massless waves should decay into two $\chi$-particles, but they cannot catch up with each other, so the remaining peaks for a massless plane wave are not observed. There is also a possibility that the ansatz used for this solution is incomplete, but this is a matter for further study.
Moreover, the interpretation through the Mathieu stability diagram cannot be properly applied to the massless case, as resonance does not emerge under these conditions. Furthermore, the ansatz employed in this solution may be incomplete, suggesting that further investigation is required to refine the theoretical framework.

    \paragraph{Acknowledgements} We thank Maxim Fitkevich, Dmitry Kirpichnikov, Dmitry Levkov, Alexander Panin and Igor Tkachev for valuable discussions.

\appendix

\section{Solution for narrow resonance regime}\label{app1}
Consider the Mathieu equation in the form
\[y''+(a-2q\cos(2z))y=0.\]
Approximation: $a\gg q$, $q>0$, q - real, a lies in a narrow unstable region where $b_m<a<a_m$ and m - peak number, $a_m, b_m$ - characteristic numbers of the Mathieu equation.

The solution for Matieu equation is in this approximation
\[y_1\simeq e^{\mu z}[C_m ce_m(z,q)+S_m se_m(z,q)],\]
\[y_2\simeq e^{-\mu z}[C_m ce_m(z,q)-S_m se_m(z,q)],\]
\[\mu\simeq\pm\frac{[(a_m-a)(a-a_m)]^{\frac{1}{2}}}{2m}\]
For the first resonant band ($a\approx1$):
\[a_1\simeq 1+q-\frac{1}{8}q^2,\]
\[b1\simeq 1-q-\frac{1}{8},\]
\[\Rightarrow \mu\simeq\pm\frac{1}{2}[(1+q-a)(a-(1-q))]^\frac{1}{2}=\frac{1}{2}\sqrt{q^2-(a-1)^2}\]
In our notations $a=A_k$:
\[\mu\simeq\frac{1}{2}\sqrt{q^2-(A_k-1)^2}.\]

\section{Condition for broad resonance}

The condition for a broad resonance is a violation of the adiabaticity condition \eqref{cond_adiab}. It follows from the fact that the WKB approximation is used in the solution \cite{lozanov2020reheating,Dufaux:2006ee}, and at the moments when it is violated and a broad resonance occurs. Then the condition for a broad resonance will be reduced to the expression
\begin{equation}\label{br_res_cond}
    \frac{d\omega}{dt}\gtrsim\omega^2.
\end{equation}
Using expression \eqref{omega_chi} for the frequency of the chi field, we obtain
\begin{equation}
    2g\dot\phi\gtrsim(k^2+m^2_\chi+2g\phi(t))^{2/3}
\end{equation}
\[\phi(t)=\Phi\cos(m_\phi t)\] 
Maximum $\phi$ when $\cos(m_\phi t)=1$ and minimum $\phi$ when $cos(m_\phi t)=0$
\[1)~\phi(t)=0 \ \mbox{for} \  m_\phi t=\pi/2+N\pi,\]
\[2)~\phi=\Phi \ \mbox{for} \  m_\phi t=2\pi N\]
Decompose the cosine into a Maclaurin series (the series converges for any argument)
\[\cos(x)=1-\frac{x^2}{2!}+...\].

1) $\dot\phi\approx-m_\phi(\frac{\pi}{2}+\pi N),$ where N is an integer.
From expression \eqref{br_res_cond} we get the condition for $k^2$ 
\begin{equation}
    k^2\lesssim(-g\Phi m_\phi(\frac{\pi}{2}+\pi N))^{2/3}-m^2_\chi-2g\phi(t).
\end{equation}
Then the condition for $\phi$ near zero will be
\begin{equation}
    \phi(t)\lesssim\frac{(-g\Phi m_\phi(\frac{\pi}{2}+\pi N))^{2/3}-m^2_\chi}{2g}.
\end{equation}

2) $\dot\phi\approx\Phi2\pi N m_\phi$, where N is an integer. From expression \eqref{br_res_cond} we get the condition for $k^2$ 
\begin{equation}
    k^2\lesssim(-g\Phi m_\phi(2\pi N))^{2/3}-m^2_\chi-2g\phi(t)
\end{equation}
Then the condition for $\phi$ near $\Phi$ will be
\[
    \phi(t)\lesssim\frac{(-g\Phi m_\phi(2\pi N))^{2/3}-m^2_\chi}{2g}.
\]

3) $\cos(m_\phi t)=-1$ when $m_\phi t=\pi+2\pi N$, then $\dot\phi\approx\Phi m_\phi(\pi+2\pi N)$
\[k^2\lesssim(-g\Phi m_\phi(\pi+2\pi N))^{2/3}-m^2_\chi-2g\phi(t)\]
and
\begin{equation}
    \phi(t)\lesssim\frac{(-g\Phi m_\phi(\pi+2\pi N))^{2/3}-m^2_\chi}{2g}.
\end{equation}

\section{Boost}\label{app_boost}

Classical plane wave of a massive scalar field is a condensate in a boosted coordinate system. Due to Lorentz invariance, the observable physical values such as the stability of the $\phi$ field configuration should remain the same in any coordinate system. In particular, the stability chart cannot depend on the plane wave frequency $\omega_{\vec p}$ or momentum $\vec p$ separately, only on the combination $\sqrt{\omega_{\vec p}^2-p^2}=m_\phi$.  The boost factor $\omega_{\vec p}/m_\phi$ describes the transfer between the laboratory and the center-of-mass systems.

To make a boost at $p\neq0$, we use Lorentz transformations \cite{landau1973field}.
Consider \[\phi'=\Phi\cos(m_\phi t)=\Phi\cos(p_\mu x_\mu),\]
where the $\phi'$ field is a scalar, respectively, the amplitude is not transformed during the Lorentz transformation.

We used the notation $'$ for the values up to the boost at $p=0$ and the parameters of the Mathieu equation in terms of the Heisenberg equation will look like this
\begin{equation}\label{A'q'}
    A'_{k'}=4\frac{k^2+m^2_\phi}{{\omega'}^2}, q=4\frac{g\Phi}{{\omega'}^2}.
\end{equation}

The density of the Hamiltonian is \[\mathcal{H}=\frac{1}{2}(\omega^2+p^2+m^2_\phi)\cos^2(p_\mu x_\mu),\]
$\omega^2-p^2=m^2_\phi$
$\mathcal{H}=\omega^2\cos^2(p_\mu x_\mu)$.
Before boost $\mathcal{H}_{m_\phi}=m^2_\phi \cos^2(m_\phi t)$,
after boost $\mathcal{H}_{p}=\omega^2_\phi \cos^2(\omega t-px)$.
We know $\mathcal{H}=T_{00}$, then when we transition from one frame of reference to another $T_{00}\rightarrow \frac{\omega^2}{m^2_\phi}T_{00}$.

Therefore, we arrive at the result that the $\gamma$factor is equal to
\[\gamma=\frac{\omega}{m_\phi}=\frac{1}{\sqrt{1-V^2}}.\]
From here we can find the velocity of the moving frame of reference
\[V=\frac{\sqrt{\omega^2-m^2_\phi}}{\omega}=\frac{p}{\omega}.\].
It follows from the Lorentz transformations for momentum that
\[k=(\pm k'+V\mathcal{E}')\gamma=(\pm k'+\frac{p}{\omega}\frac{m_\phi}{2})\frac{\omega}{m_\phi}=\pm k'\frac{\omega}{m_\phi}+\frac{p}{2},\]
is where $\mathcal{E}'=\frac{m_\phi}{2}$  is the energy of the resulting particle in the reference frame with momentum $p=0$ and if we consider the boundary case $k'=0$, then as a result of the boost $k$ will be equal to $p/2$. It is this boundary case that we consider in the following paragraphs.

After the boost, we can no longer apply Mathieu's methods to solving the equation. However, we can study how the relationship between the parameters of the Mathieu equation $A_k$ and $q$ will change after the boost $p\neq0$.
\[A_k=\frac{4k^2}{\omega^2}+\frac{4m^2_\chi}{\omega^2}, \quad q=\frac{4g\Phi}{\omega^2},\] 
where $\omega^2=m^2_{\phi}+p^2$.

\bibliography{biblio}
\end{document}